\documentclass{article}
\usepackage{amssymb}
\usepackage{amsmath}
\usepackage{amsfonts}
\usepackage{sw20ssur}
\usepackage{cite}
\usepackage{afterpage}

\input{tcilatex}
\begin{document}

\title{{\LARGE A probabilistic approach to quantum Bayesian games of
incomplete information}}
\author{Azhar Iqbal$^{\dagger \ddagger }$, James M. Chappell$^{^{\ddagger }}$%
, Qiang Li$^{\diamond }$, Charles E.~M.~Pearce$^{\bigtriangleup }$, and
Derek Abbott$^{^{\ddagger }}$ \\
$^{\dagger }${\small Department of Mathematics \& Statistics, King Fahd
University of Petroleum \&}\\
{\small \ Minerals, Dhahran 31261, Kingdom of Saudi Arabia.}\\
$^{\ddagger }${\small School of Electrical \& Electronic Engineering, The
University of Adelaide, SA 5005, Australia.}\\
$^{\diamond }${\small College of Electrical Engineering, Chongqing
University, Chongqing 400030, People's Republic of China.}\\
$^{\bigtriangleup }${\small School of Mathematical Sciences, The University
of Adelaide, SA 5005, Australia.}}
\maketitle

\begin{abstract}
A Bayesian game is a game of incomplete information in which the rules of
the game are not fully known to all players. We consider the Bayesian game
of Battle of Sexes that has several Bayesian Nash equilibria and investigate
its outcome when the underlying probability set is obtained from generalized
Einstein-Podolsky-Rosen experiments. We find that this probability set,
which may become non-factorizable, results in a unique Bayesian Nash
equilibrium of the game.
\end{abstract}

\section{Introduction}

The standard approach to constructing quantum games \cite%
{Blaquiere,Wiesner,Mermin,Mermin1,MeyerDavid,EWL,Vaidman,BenjaminHayden,EnkPike,Johnson,MarinattoWeber,IqbalToor1,DuLi,Du,Piotrowski,IqbalToor3,FlitneyAbbott1,IqbalToor2,Piotrowski1,Shimamura1, FlitneyAbbott2,DuNplayer,Hanetal,IqbalWeigert,Mendes,CheonTsutsui,IqbalEPR,NawazToor,OzdemirA,Cheon,Shimamura,ChenWang,Schmidetal,IchikawaTsutsui,CheonIqbal,Ozdemir,FlitneyGreentree,IqbalCheon,IchikawaTsutsuiCheon,Ramzan,FlitneyHollenberg, Aharon,Bleiler,AhmedBleilerKhan,Qiang,Qiang1,ChappellA,ChappellD,IqbalAbbott,Chappell,IqbalCheonAbbott,ChappellB,ChappellC,Phoenix,Khan1,Khan2}
naturally uses the formalism of quantum mechanics in Hilbert space \cite%
{Peres}. In recent years, however, a probabilistic approach to this research
area \cite%
{IqbalEPR,IqbalCheon,IqbalCheonAbbott,IqbalAbbott,ChappellC,ChappellD,ChappellB,ChappellA}
has been proposed that uses sets of non-factorizable quantum mechanical
probabilities, i.e.~without using the quantum mechanical concepts of state
vectors, self-adjoint operators, and quantum measurement etc. This is with a
view of making the area of quantum games more accessible to wider
mathematical application, as the methods and range of solution concepts of
game theory \cite{vonNeumann,Binmore,Rasmusen,Osborne} are used and
exploited, without any real need for invoking the Hilbert space formalism of
quantum mechanics.

The probabilistic approach for a two-player two-strategy game directly uses
sets of quantum probabilities corresponding to the measurement outcomes on a
two qubit quantum system. As is known, the setting of generalized
Einstein-Podolsky-Rosen (EPR) experiments \cite%
{EPR,Bohm,Bell,Bell1,Bell2,Aspect,ClauserShimony,CHSH,Cereceda} performed on
this system leads to the consideration of a set of $16$ quantum
probabilities. Properties of this probability set have been investigated by
Cereceda \cite{Cereceda} and it has been pointed out that the CHSH form of
Bell's inequality \cite{CHSH} can be re-expressed in terms of constraints on
the elements from this set. It is observed that only a non-factorizable
probability set, as is defined later, can lead to the violation of Bell's
inequality and that not every non-factorizable probability set violates
Bell's inequality.

A Bayesian game is a game of incomplete information in which the rules of
the game are not fully known to all players. In this paper, we study a
Bayesian game that is a variant of the well known Battle of Sexes game, also
studied \cite{MarinattoWeber,CheonIqbal} in the quantum game literature. In
an earlier study \cite{CheonIqbal}, using the quantization protocol based on
Schmidt decomposition \cite{Peres}, a Bayesian game of incomplete
information \cite{Osborne} has been investigated in relation to the
violation of Bell's inequality \cite{Bell,Bell1,Bell2,Aspect,Peres}. The
present paper, however, adopts a different approach in that, without
referring to the Hilbert space formalism, it finds the outcome of a Bayesian
game when the considered probability set can be non-factorizable---this can
arise, in an experimental situation, from a set up involving quantum
entanglement, such as an EPR-type experiment, where the CHSH form of Bell's
inequality is violated.

In its normal form representation, the game matrix of the Bayesian game has
the same number of entries as the $16$ elements $\varepsilon _{i}$ in the
probability set that corresponds to the generalized EPR experiments \cite%
{Cereceda}. We find that the richer structure of the Bayesian game permits a
natural embedding of the classical factorizable game within the quantum
game. We show that, whereas the classical factorizable Bayesian game of
imperfect information has several Nash equilibria, its non-factorizable
quantum version obtained from a set of quantum probabilities corresponding
to generalized EPR experiments has a unique Nash equilibrium.

The suggested probabilistic approach to obtaining quantum games thus
re-expresses players' payoff relations in terms of a set of probabilities
that can also arise in a quantum mechanical experiment. As the approach is
based on probabilities only, it does not refer to the formalism of quantum
mechanics using state vectors, unitary transformations, and quantum
measurements etc. As game theory is a broad area, with applications ranging
from trade, politics, sociology, biology, engineering etc., most researchers
in this area are naturally not familiar with the mathematical formalism of
quantum mechanics. This paper thus fills in that gap and demonstrates how an
unusual game-theoretic outcome for a Bayesian game results when probability
sets that are obtained in quantum mechanical experiments are the underlying
probabilities of a Bayesian game. In this approach, the quantum game reduces
itself to the classical game when the considered probability set becomes
factorizable.

The rest of this paper is organized as follows. Sections~2-3 present a
review of the classical theory of a Bayesian game that is a variant of the
game of Battle of Sexes. Section~4 describes quantum probabilities in
generalized EPR experiments, their constraints, and how within the quantum
game the players' payoff relations are defined in terms of these
probabilities. Section~5 analyses the outcome of the Bayesian game of Battle
of Sexes with EPR probabilities and Section~6 discusses the results.

\section{The Bayesian game of Battle of Sexes and its variant}

The game of Battle of Sexes (BoS) describes \cite{Osborne} the following
situation. Two people Alice and Bob wish to go out together and two concerts
are available: one with music by Bach, and one with music by Stravinsky. One
person prefers Bach and the other prefers Stravinsky. If they go to
different concerts, each of them is equally unhappy listening to the music
of either composer. The situation is represented by the following matrix

\begin{equation}
\begin{array}{c}
\text{Alice}%
\end{array}%
\begin{array}{c}
\mathcal{B} \\ 
\mathcal{S}%
\end{array}%
\overset{\overset{%
\begin{array}{c}
\text{Bob}%
\end{array}%
}{%
\begin{array}{ccc}
\mathcal{B} &  & \mathcal{S}%
\end{array}%
}}{\left( 
\begin{array}{cc}
(2,1) & (0,0) \\ 
(0,0) & (1,2)%
\end{array}%
\right) },
\end{equation}%
where Bach and Stravinsky are represented by symbols $\mathcal{B}$ and $%
\mathcal{S}$, respectively. For this game, a Nash equilibrium (NE) is a pair
of strategies such that each player's strategy is the best reply to the
strategic choice of the other players. In other words, unilateral deviation
from a Nash equilibrium by a player in the form of a different choice of
strategy will produce a payoff that is less than or equal to what a Nash
equilibrium strategy will give to that player. Analysis shows \cite{Osborne}
that this game has three mixed strategy Nash equilibria $\left( 0,0\right) ,$
$(\frac{2}{3},\frac{1}{3}),$ $(1,1)$, where the numbers in parentheses are
the Alice's and Bob's probabilities of choosing the strategy $\mathcal{B}$.

An interesting variant of this game \cite{Osborne} is the one in which Alice
is unsure whether Bob prefers to join her or prefers to avoid her, whereas
Bob knows Alice's preferences. Assume that Alice thinks that with
probability $\frac{1}{2}$ Bob wants to go out with her, and with probability 
$\frac{1}{2}$ Bob wants to avoid her,

\begin{equation}
\overset{%
\begin{array}{c}
\text{Alice playing against Bob's two types}%
\end{array}%
}{\overset{%
\begin{array}{c}
\text{probability }\frac{1}{2}%
\end{array}%
}{\underset{%
\begin{array}{c}
\text{Bob's first type}%
\end{array}%
}{%
\begin{array}{c}
\mathcal{B} \\ 
\mathcal{S}%
\end{array}%
\overset{%
\begin{array}{ccc}
\mathcal{B} &  & \mathcal{S}%
\end{array}%
}{\left( 
\begin{array}{cc}
(2,1) & (0,0) \\ 
(0,0) & (1,2)%
\end{array}%
\right) }}}\text{ \ \ \ \ \ \ \ \ \ \ }\overset{%
\begin{array}{c}
\text{probability }\frac{1}{2}%
\end{array}%
}{\underset{%
\begin{array}{c}
\text{Bob's second type}%
\end{array}%
}{%
\begin{array}{c}
\mathcal{B} \\ 
\mathcal{S}%
\end{array}%
\overset{%
\begin{array}{ccc}
\mathcal{B} &  & \mathcal{S}%
\end{array}%
}{\left( 
\begin{array}{cc}
(2,0) & (0,2) \\ 
(0,1) & (1,0)%
\end{array}%
\right) .}}}}  \label{Bob's two types}
\end{equation}%
That is, from Alice's perspective, Bob has two possible types, the first is
shown on the left and the second is on the right in (\ref{Bob's two types}).

Alice does not know Bob's type and is thus faced with the situation of
choosing her rational action that is based on her belief about the action of
Bob of each type. Given these beliefs, and her belief about the likelihood
of each type, she can calculate her expected payoff in each case. For
instance, given that Alice plays $\mathcal{B}$, and Bob of first type (who
wishes to meet Alice) plays $\mathcal{B}$ whereas Bob of second type (who
wishes to avoid Alice) plays $\mathcal{S}$, then Alice's expected payoff is $%
\frac{1}{2}(2)+\frac{1}{2}(0)=1$.

There are $4$ possible pairs of actions of Bob's two types given as $(%
\mathcal{B},\mathcal{B}),$ $(\mathcal{B},\mathcal{S}),$ $(\mathcal{S},%
\mathcal{B}),$ and $(\mathcal{S},\mathcal{S})$. Here, for instance, $(%
\mathcal{S},\mathcal{B})$ describes that Bob's first type plays $\mathcal{S}$
and Bob's second type plays $\mathcal{B}$. Players' expected payoffs are
then obtained as given below.

\begin{equation}
\begin{array}{c}
\text{Alice}%
\end{array}%
\overset{%
\begin{array}{c}
\text{Bob's two types}%
\end{array}%
}{%
\begin{array}{cccccccc}
& (\mathcal{B},\mathcal{B}) &  & (\mathcal{B},\mathcal{S}) &  & (\mathcal{S},%
\mathcal{B}) &  & (\mathcal{S},\mathcal{S}) \\ 
\mathcal{B} & (2,\frac{1}{2}) &  & (1,\frac{3}{2}) &  & (1,0) &  & (0,1) \\ 
\mathcal{S} & (0,\frac{1}{2}) &  & (\frac{1}{2},0) &  & (\frac{1}{2},\frac{3%
}{2}) &  & (1,1)%
\end{array}%
}.  \label{Table 1}
\end{equation}

One can then show \cite{Osborne} that the triplet $(\mathcal{B},(\mathcal{B},%
\mathcal{S}))$, where the first entry in the bracket is Alice's action $%
\mathcal{B}$ and $(\mathcal{B},\mathcal{S})$ is the pair of actions of the
two types of Bob, constitutes a Nash equilibrium. This is a pure strategy
Nash equilibrium consisting of three actions, one for Alice and one for each
of the two types of Bob, with the property that a) Alice's action is optimal
given the actions of the two types of Bob, b) the action of each type of Bob
is optimal given the action of Alice.

\section{Battle of Sexes with imperfect information}

We now consider a situation in which neither player knows whether the other
wants to meet or not \cite{Osborne}. As before, Alice assumes that Bob will
prefer to join her with probability $\frac{1}{2}$ and will prefer to avoid
her with probability $\frac{1}{2}$. Moreover, Bob expects with probability $%
\omega $ Alice will prefer to join him and with probability $(1-\omega )$
that she will prefer to avoid him. It is assumed that both Alice and Bob
know their own preferences. The game can be represented as shown in Fig.~1 
\cite{Osborne}.

\FRAME{ftbpFU}{4.8922in}{5.6239in}{0pt}{\Qcb{The game of Battle of Sexes
when each player is unsure of other player's preferences \protect\cite%
{Osborne}. }}{\Qlb{Bayesian}}{bayesian.ps}{\special{language "Scientific
Word";type "GRAPHIC";maintain-aspect-ratio TRUE;display "USEDEF";valid_file
"F";width 4.8922in;height 5.6239in;depth 0pt;original-width
8.2685in;original-height 11.6949in;cropleft "0.0123";croptop
"0.9409";cropright "0.9766";cropbottom "0.1564";filename
'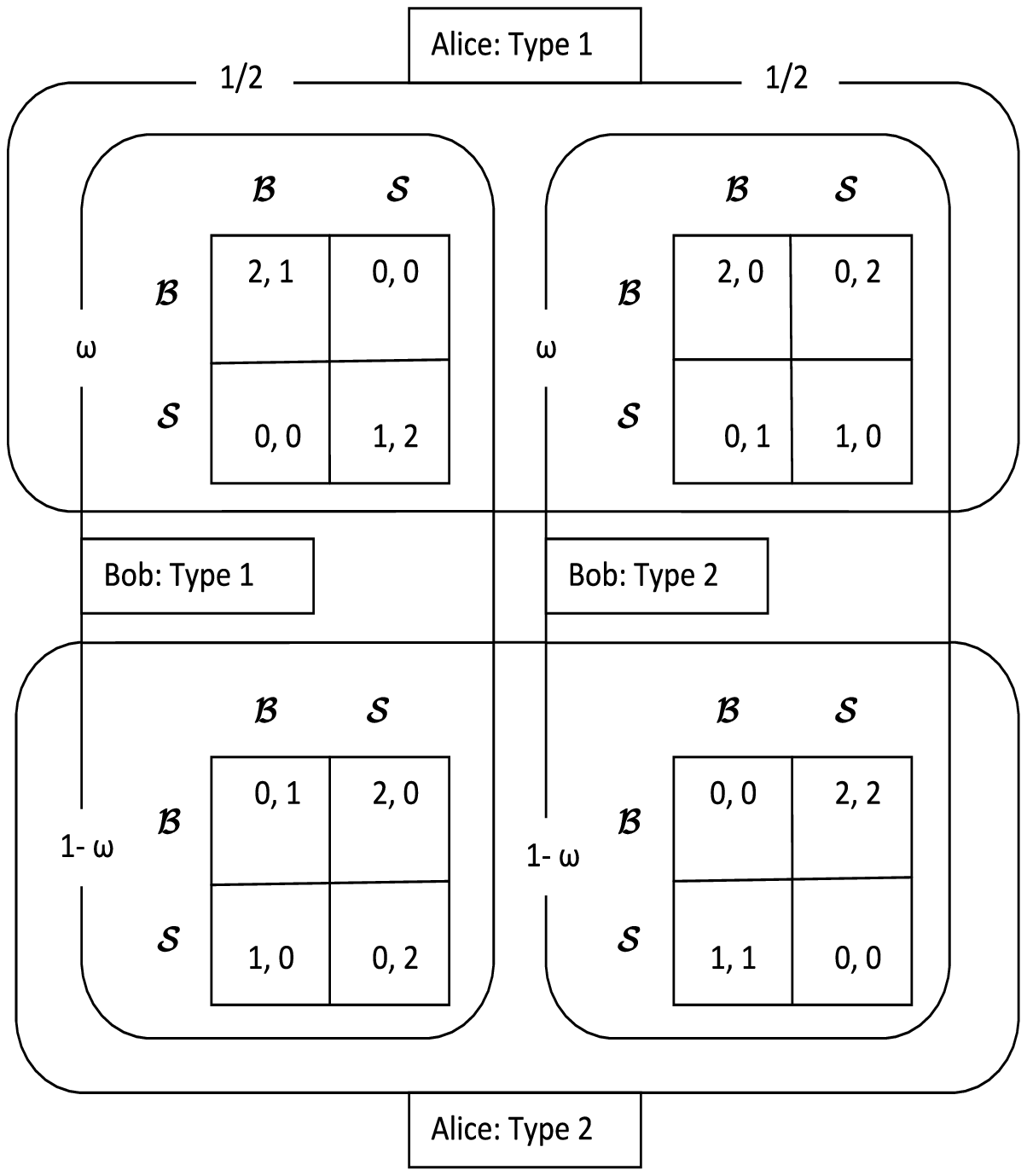';file-properties "XNPEU";}}

Consider the payoffs of Alice of type $1$. She believes that with
probability $\frac{1}{2}$ she faces Bob of type $1$ and with probability $%
\frac{1}{2}$ she faces Bob of type $2$. Assume that Bob of type $1$ chooses $%
\mathcal{B}$ and Bob of type $2$ chooses $\mathcal{S}$. Then if Alice of
type $1$ chooses $\mathcal{B}$, her expected payoff is $\frac{1}{2}(2)+\frac{%
1}{2}(0)=1$ and if she chooses $\mathcal{S}$, her expected payoff is $\frac{1%
}{2}(0)+\frac{1}{2}(1)=\frac{1}{2}$.

We represent the action of the two types of Bob by the pairs $(\mathcal{B},%
\mathcal{B}),$ $(\mathcal{B},\mathcal{S}),$ $(\mathcal{S},\mathcal{B}),$ $(%
\mathcal{S},\mathcal{S})$, where the first entry in the bracket is the
action of Bob of type $1$ and the second entry is the action of Bob of type $%
2$. The expected payoff to Alice of type $1$ when she chooses $\mathcal{B}$
or $\mathcal{S}$, against the $4$ pairs of actions of the two types of Bob
are

\begin{equation}
\begin{array}{c}
\text{Alice of type }1%
\end{array}%
\overset{%
\begin{array}{c}
\text{Bob's two types}%
\end{array}%
}{%
\begin{array}{cccccccc}
& (\mathcal{B},\mathcal{B}) &  & (\mathcal{B},\mathcal{S}) &  & (\mathcal{S},%
\mathcal{B}) &  & (\mathcal{S},\mathcal{S}) \\ 
\mathcal{B} & 2 &  & 1 &  & 1 &  & 0 \\ 
\mathcal{S} & 0 &  & \frac{1}{2} &  & \frac{1}{2} &  & 1%
\end{array}%
},  \label{AliceVSBob1}
\end{equation}%
and similarly the payoff to Alice of type $2$ is found as

\begin{equation}
\begin{array}{c}
\text{Alice of type }2%
\end{array}%
\overset{%
\begin{array}{c}
\text{Bob's two types}%
\end{array}%
}{%
\begin{array}{cccccccc}
& (\mathcal{B},\mathcal{B}) &  & (\mathcal{B},\mathcal{S}) &  & (\mathcal{S},%
\mathcal{B}) &  & (\mathcal{S},\mathcal{S}) \\ 
\mathcal{B} & 0 &  & 1 &  & 1 &  & 2 \\ 
\mathcal{S} & 1 &  & \frac{1}{2} &  & \frac{1}{2} &  & 0%
\end{array}%
.}  \label{AliceVSBob2}
\end{equation}

Consider the case when Alice of type $1$ plays $\mathcal{B}$, Alice of type $%
2$ plays $\mathcal{S}$, Bob of type $1$ plays $\mathcal{S}$, and Bob of type 
$2$ plays $\mathcal{B}$. We represent this case by the quadruple $(\mathcal{B%
},\mathcal{S}),(\mathcal{S},\mathcal{B})$ where the entries in the first
pair are chosen by Alice's two types, respectively, and the entries in the
second pair are chosen by Bob's two types, respectively. For this quadruple,
the payoffs to Alice's two types can be found from (\ref{AliceVSBob1},\ref%
{AliceVSBob2}) at the entries located at the intersection of the same column
with entry $(\mathcal{S},\mathcal{B})$ and the two row entries at $\mathcal{B%
}$ and $\mathcal{S}$ corresponding to Alice of type $1$ and type $2$,
respectively:

\begin{equation}
\Pi _{\text{A}_{1}}\left\{ (\mathcal{B},\mathcal{S}),(\mathcal{S},\mathcal{B}%
)\right\} =1,\text{ }\Pi _{\text{A}_{2}}\left\{ (\mathcal{B},\mathcal{S}),(%
\mathcal{S},\mathcal{B})\right\} =\frac{1}{2},
\end{equation}%
where the subscripts $1$ \& $2$ for A or B give the type of that player.

The following table gives the expected payoffs to Alice's two types against
the pairs of actions $(\mathcal{B},\mathcal{B}),(\mathcal{B},\mathcal{S}),(%
\mathcal{S},\mathcal{B}),(\mathcal{S},\mathcal{S})$ by Bob's two types,
respectively

\begin{equation}
\begin{array}{c}
\text{Alice's two types}%
\end{array}%
\overset{%
\begin{array}{c}
\text{Bob's two types}%
\end{array}%
}{%
\begin{array}{cccccccc}
& (\text{$\mathcal{B}$},\text{$\mathcal{B}$}) &  & (\text{$\mathcal{B}$},%
\mathcal{S}) &  & (\mathcal{S},\text{$\mathcal{B}$}) &  & (\mathcal{S},%
\mathcal{S}) \\ 
(\mathcal{B},\text{$\mathcal{B}$}) & (2,0) &  & (1,1) &  & (1,1) &  & (0,2)
\\ 
(\text{$\mathcal{B}$},\mathcal{S}) & (2,1) &  & (1,\frac{1}{2}) &  & (1,%
\frac{1}{2}) &  & (0,0) \\ 
(\mathcal{S},\text{$\mathcal{B}$}) & (0,0) &  & (\frac{1}{2},1) &  & (\frac{1%
}{2},1) &  & (1,2) \\ 
(\mathcal{S},\mathcal{S}) & (0,1) &  & (\frac{1}{2},\frac{1}{2}) &  & (\frac{%
1}{2},\frac{1}{2}) &  & (1,0)%
\end{array}%
},  \label{Alice's table}
\end{equation}%
where in the column on the left, for instance, $(\mathcal{B},\mathcal{S})$
means that Alice of type $1$ chooses $\mathcal{B}$ and Alice of type $2$
chooses $\mathcal{S}$. The two payoff entries in brackets are the expected
payoffs to Alice of type $1$ and type $2$, respectively. In (\ref{Alice's
table}) consider the entry $(\frac{1}{2},1)$ at the intersection of $3$rd
row $(\mathcal{S},\mathcal{B})$ and $3$rd column $(\mathcal{S},\mathcal{B})$%
. This means that Alice of the types $1$ and $2$ chooses $\mathcal{S}$ and $%
\mathcal{B}$, respectively, and Bob of the types $1$ and $2$ also chooses $%
\mathcal{S}$ and $\mathcal{B}$, respectively. The payoff to Alice of type $1$
and type $2$ are $\frac{1}{2}$ and $1$, respectively.

Similarly, the payoffs to Bob's two types can be found as

\begin{eqnarray}
\Pi _{\text{B}_{1}}\left\{ (\mathcal{B},\mathcal{S}),(\mathcal{S},\mathcal{B}%
)\right\} &=&\omega (0)+(1-\omega )(2)=2(1-\omega ), \\
\Pi _{\text{B}_{2}}\left\{ (\mathcal{B},\mathcal{S}),(\mathcal{S},\mathcal{B}%
)\right\} &=&\omega (0)+(1-\omega )(1)=1-\omega ,
\end{eqnarray}%
and one can write the expected payoffs to Bob's first type as

\begin{equation}
\begin{array}{c}
\text{Bob's type }1%
\end{array}%
\overset{%
\begin{array}{c}
\text{Alice's two types}%
\end{array}%
}{%
\begin{array}{cccccccc}
& (\mathcal{B},\mathcal{B}) &  & (\mathcal{B},\mathcal{S}) &  & (\mathcal{S},%
\mathcal{B}) &  & (\mathcal{S},\mathcal{S}) \\ 
\mathcal{B} & 1 &  & \omega &  & 1-\omega &  & 0 \\ 
\mathcal{S} & 0 &  & 2(1-\omega ) &  & 2\omega &  & 2%
\end{array}%
},  \label{xyz1}
\end{equation}%
and, similarly, the expected payoffs to Bob's second type are obtained as

\begin{equation}
\begin{array}{c}
\text{Bob's type }2%
\end{array}%
\overset{%
\begin{array}{c}
\text{Alice's two types}%
\end{array}%
}{%
\begin{array}{cccccccc}
& (\mathcal{B},\mathcal{B}) &  & (\mathcal{B},\mathcal{S}) &  & (\mathcal{S},%
\mathcal{B}) &  & (\mathcal{S},\mathcal{S}) \\ 
\mathcal{B} & 0 &  & 1-\omega &  & \omega &  & 1 \\ 
\mathcal{S} & 2 &  & 2\omega &  & 2(1-\omega ) &  & 0%
\end{array}%
}.  \label{xyz2}
\end{equation}%
As it is the case for payoffs to Alice's two types above, (\ref{xyz1}) and (%
\ref{xyz2}) can be joined together to obtain the payoffs to Bob's two types
as

\begin{equation}
\begin{array}{c}
\text{Alice's two types}%
\end{array}%
\overset{%
\begin{array}{c}
\text{Bob's two types}%
\end{array}%
}{%
\begin{array}{cccccccc}
& \text{$(\mathcal{B},\mathcal{B})$} &  & \text{$(\mathcal{B},\mathcal{S})$}
&  & \text{$(\mathcal{S},\mathcal{B})$} &  & \text{$(\mathcal{S},\mathcal{S}%
) $} \\ 
\text{$(\mathcal{B},\mathcal{B})$} & (1,0) &  & (1,2) &  & (0,0) &  & (0,2)
\\ 
\text{$(\mathcal{B},\mathcal{S})$} & (\omega ,1-\omega ) &  & (\omega
,2\omega ) &  & (2(1-\omega ),(1-\omega )) &  & (2(1-\omega ),2\omega ) \\ 
\text{$(\mathcal{S},\mathcal{B})$} & ((1-\omega ),\omega ) &  & ((1-\omega
),2(1-\omega )) &  & (2\omega ,\omega ) &  & (2\omega ,2(1-\omega )) \\ 
\text{$(\mathcal{S},\mathcal{S})$} & (0,1) &  & (0,0) &  & (2,1) &  & (2,0)%
\end{array}%
},  \label{Bob's table}
\end{equation}%
where the entries in brackets are the expected payoffs to Bob of type $1$
and type $2$, respectively. Now, (\ref{Alice's table}) and (\ref{Bob's table}%
) can be joined together to obtain

\begin{align}
& 
\begin{array}{c}
\text{Alice}%
\end{array}%
\overset{%
\begin{array}{c}
\text{Bob}%
\end{array}%
}{%
\begin{array}{cccccccc}
& \text{{\small $(\mathcal{B},\mathcal{B})$}} &  & \text{{\small $(\mathcal{B%
},\mathcal{S})$}} &  & \text{{\small $(\mathcal{S},\mathcal{B})$}} &  & 
\text{{\small $(\mathcal{S},\mathcal{S})$}} \\ 
\text{{\small $(\mathcal{B},\mathcal{B})$}} & {\small (2,0),(1,0)} &  & 
{\small (1,1),(1,2)} &  & {\small (1,1),(0,0)} &  & {\small (0,2),(0,2)} \\ 
\text{{\small $(\mathcal{B},\mathcal{S})$}} & {\small (2,1),(\omega
,1-\omega )} &  & {\small (1,}\frac{1}{2}{\small ),(\omega ,2\omega )} &  & 
{\small (1,}\frac{1}{2}{\small ),(2(1-\omega ),(1-\omega ))} &  & {\small %
(0,0),(2(1-\omega ),2\omega )} \\ 
\text{{\small $(\mathcal{S},\mathcal{B})$}} & {\small (0,0),((1-\omega
),\omega )} &  & {\small (}\frac{1}{2}{\small ,1),((1-\omega ),2(1-\omega ))}
&  & {\small (}\frac{1}{2}{\small ,1),(2\omega ,\omega )} &  & {\small %
(1,2),(2\omega ,2(1-\omega ))} \\ 
\text{{\small $(\mathcal{S},\mathcal{S})$}} & {\small (0,1),(0,1)} &  & 
{\small (}\frac{1}{2}{\small ,}\frac{1}{2}{\small ),(0,0)} &  & {\small (}%
\frac{1}{2}{\small ,}\frac{1}{2}{\small ),(2,1)} &  & {\small (1,0),(2,0)}%
\end{array}%
}  \notag \\
&
\end{align}%
where, for the two pairs of payoff entries, the first pair is for Alice's
two types and the second payoff pair is for Bob's two types. It can be seen
that when $\omega =\frac{2}{3}$, for instance, the strategy quadruples $%
\left\{ (\mathcal{S},\mathcal{B}),(\mathcal{S},\mathcal{S})\right\} $ and $%
\left\{ (\mathcal{B},\mathcal{B}),(\mathcal{B},\mathcal{S})\right\} $
corresponding to the payoff quadruples $(1,2),(\frac{4}{3},\frac{2}{3})$ and 
$(1,1),(1,2)$, respectively, are the pure strategy Bayesian Nash equilibria 
\cite{Rasmusen,Osborne}.

\subsection{Mixed-strategy version}

Now consider the mixed-strategy version of the game in which the players'
probabilities of selecting $\mathcal{B}$ from the pure strategies $\mathcal{B%
}$ \& $\mathcal{S}$ are given by numbers $p,q,p^{\prime },$ and $q^{\prime }$
$\in \left[ 0,1\right] $ for Alice of type $1$, Alice of type $2$, Bob of
type $1$, and for Bob of type $2$, respectively. The mixed-strategy payoffs
for Alice's and Bob's two types can then be found from Fig.~1 as

\begin{align}
\Pi _{\text{A}_{1}}(p;p^{\prime },q^{\prime })& =\frac{1}{2}\left( 
\begin{array}{c}
p \\ 
1-p%
\end{array}%
\right) ^{T}\left( 
\begin{array}{cc}
2 & 0 \\ 
0 & 1%
\end{array}%
\right) \left( 
\begin{array}{c}
p^{\prime } \\ 
1-p^{\prime }%
\end{array}%
\right) +\frac{1}{2}\left( 
\begin{array}{c}
p \\ 
1-p%
\end{array}%
\right) ^{T}\left( 
\begin{array}{cc}
2 & 0 \\ 
0 & 1%
\end{array}%
\right) \left( 
\begin{array}{c}
q^{\prime } \\ 
1-q^{\prime }%
\end{array}%
\right) ,  \notag \\
\Pi _{\text{A}_{2}}(q;p^{\prime },q^{\prime })& =\frac{1}{2}\left( 
\begin{array}{c}
q \\ 
1-q%
\end{array}%
\right) ^{T}\left( 
\begin{array}{cc}
0 & 2 \\ 
1 & 0%
\end{array}%
\right) \left( 
\begin{array}{c}
p^{\prime } \\ 
1-p^{\prime }%
\end{array}%
\right) +\frac{1}{2}\left( 
\begin{array}{c}
q \\ 
1-q%
\end{array}%
\right) ^{T}\left( 
\begin{array}{cc}
0 & 2 \\ 
1 & 0%
\end{array}%
\right) \left( 
\begin{array}{c}
q^{\prime } \\ 
1-q^{\prime }%
\end{array}%
\right) ,  \notag \\
\Pi _{\text{B}_{1}}(p^{\prime };p,q)& =\omega \left( 
\begin{array}{c}
p \\ 
1-p%
\end{array}%
\right) ^{T}\left( 
\begin{array}{cc}
1 & 0 \\ 
0 & 2%
\end{array}%
\right) \left( 
\begin{array}{c}
p^{\prime } \\ 
1-p^{\prime }%
\end{array}%
\right) +(1-\omega )\left( 
\begin{array}{c}
q \\ 
1-q%
\end{array}%
\right) ^{T}\left( 
\begin{array}{cc}
1 & 0 \\ 
0 & 2%
\end{array}%
\right) \left( 
\begin{array}{c}
p^{\prime } \\ 
1-p^{\prime }%
\end{array}%
\right) ,  \notag \\
\Pi _{\text{B}_{2}}(q^{\prime };p,q)& =\omega \left( 
\begin{array}{c}
p \\ 
1-p%
\end{array}%
\right) ^{T}\left( 
\begin{array}{cc}
0 & 2 \\ 
1 & 0%
\end{array}%
\right) \left( 
\begin{array}{c}
q^{\prime } \\ 
1-q^{\prime }%
\end{array}%
\right) +(1-\omega )\left( 
\begin{array}{c}
q \\ 
1-q%
\end{array}%
\right) ^{T}\left( 
\begin{array}{cc}
0 & 2 \\ 
1 & 0%
\end{array}%
\right) \left( 
\begin{array}{c}
q^{\prime } \\ 
1-q^{\prime }%
\end{array}%
\right) ,  \notag \\
&  \label{MixedStrategyPayoffs}
\end{align}%
where $T$ is for transpose and the subscripts $1$ and $2$ under A and B
refer to the respective player's type. Also, a semicolon is used to separate
Alice's and Bob's variables. The following Nash inequalities are then
obtained

\begin{eqnarray}
\Pi _{\text{A}_{1}}(p^{\ast };p^{\prime \ast },q^{\prime \ast })-\Pi _{\text{%
A}_{1}}(p,p^{\prime \ast },q^{\prime \ast }) &=&\frac{\partial \Pi _{\text{A}%
_{1}}}{\partial p}\mid _{\ast }(p^{\ast }-p)=\left\{ \frac{3}{2}(p^{\prime
\ast }+q^{\prime \ast })-1\right\} (p^{\ast }-p)\geq 0,  \notag \\
\Pi _{\text{A}_{2}}(q^{\ast };p^{\prime \ast },q^{\prime \ast })-\Pi _{\text{%
A}_{2}}(q,p^{\prime \ast },q^{\prime \ast }) &=&\frac{\partial \Pi _{\text{A}%
_{2}}}{\partial q}\mid _{\ast }(q^{\ast }-q)=2\left\{ 1-(p^{\prime \ast
}+q^{\prime \ast })\right\} (q^{\ast }-p)\geq 0,  \notag \\
\Pi _{\text{B}_{1}}(p^{\prime \ast };p^{\ast },q^{\ast })-\Pi _{\text{B}%
_{1}}(p^{\prime };p^{\ast },q^{\ast }) &=&\frac{\partial \Pi _{\text{B}_{1}}%
}{\partial p^{\prime }}\mid _{\ast }(p^{\prime \ast }-p^{\prime })=\left\{
3\omega p^{\ast }+3(1-\omega )q^{\ast }-2\right\} (p^{\prime \ast
}-p^{\prime })\geq 0,  \notag \\
\Pi _{\text{B}_{2}}(q^{\prime \ast };p^{\ast },q^{\ast })-\Pi _{\text{B}%
_{2}}(q^{\prime };p^{\ast },q^{\ast }) &=&\frac{\partial \Pi _{\text{B}_{2}}%
}{\partial q^{\prime }}\mid _{\ast }(q^{\prime \ast }-q^{\prime })=-\left\{
3\omega p^{\ast }+3(1-\omega )q^{\ast }-1\right\} (q^{\prime \ast
}-q^{\prime })\geq 0,  \notag \\
&&  \label{NashEquilibria}
\end{eqnarray}%
where the quadruple $\left\{ (p^{\ast },q^{\ast }),(p^{\prime \ast
},q^{\prime \ast })\right\} $ is the Nash equilibrium strategy set, which is
indicated by use of the asterisk label. From the inequalities (\ref%
{NashEquilibria}), at $\omega =\frac{2}{3}$, for instance, the pure Bayesian
Nash equilibria quadruples can be identified as $\left\{ (0,1),(0,0)\right\} 
$ and $\left\{ (1,1),(1,0)\right\} $, which correspond to the strategy
quadruples $\left\{ (\mathcal{S},\mathcal{B}),(\mathcal{S},\mathcal{S}%
)\right\} $ and $\left\{ (\mathcal{B},\mathcal{B}),(\mathcal{B},\mathcal{S}%
)\right\} $ as can be observed above. Also, it is observed that $\left\{ (%
\frac{1}{2},1),(\frac{2}{3},0)\right\} $, for instance, is a mixed-strategy
Bayesian Nash equilibrium at which the players' payoffs are obtained from
Eqs.~(\ref{MixedStrategyPayoffs}) as

\begin{eqnarray}
\Pi _{\text{A}_{1}}(\frac{1}{2};\frac{2}{3},0) &=&\frac{2}{3},\text{ \ \ }%
\Pi _{\text{A}_{2}}(1;\frac{2}{3},0)=\frac{4}{3},  \notag \\
\Pi _{\text{B}_{1}}(\frac{2}{3};\frac{1}{2},1) &=&\frac{2}{3},\text{ \ \ }%
\Pi _{\text{B}_{2}}(0;\frac{1}{2},1)=2-\omega .
\end{eqnarray}

\section{Quantum probabilities in generalized Einstein-Podolsky-Rosen
experiments}

The above analysis of mixed-strategy Bayesian Nash equilibria assumes the
underlying probabilities to be factorizable. We now would like to know
whether the outcome of the Bayesian game from Fig.~1 is affected when the
probabilities become non-factorizable. For this we would consider the set of
non-factorizable quantum probabilities obtained from generalized EPR
experiments. The standard setting of such experiments \cite%
{EPR,Bohm,Bell,Bell1,Bell2,Aspect,ClauserShimony,CHSH,Peres,Cereceda}
involves a large number of runs. Two halves of an EPR pair originate from
the same source travelling in opposite directions. One half is received by
observer 1 whereas observer 2 receives the other half. The two observers are
space like separated and are unable to communicate.

\FRAME{ftbpFU}{3.4169in}{1.9138in}{0pt}{\Qcb{Figure 2: The setting for
generalized Einstein-Podolsky-Rosen experiments. We associate Alice's two
directions to the two types of Alice i.e. $D_{1}\sim $Alice's type 1 and $%
D_{2}\sim $Alice's type $2$. Similarly, $D_{1}^{\prime }\sim $Bob's type 1
and $D_{2}^{\prime }\sim $Bob's type 2.}}{\Qlb{Fig2}}{epr.ps}{\special%
{language "Scientific Word";type "GRAPHIC";maintain-aspect-ratio
TRUE;display "USEDEF";valid_file "F";width 3.4169in;height 1.9138in;depth
0pt;original-width 8.2685in;original-height 11.6949in;cropleft
"0.2756";croptop "0.6660";cropright "0.7072";cropbottom "0.4962";filename
'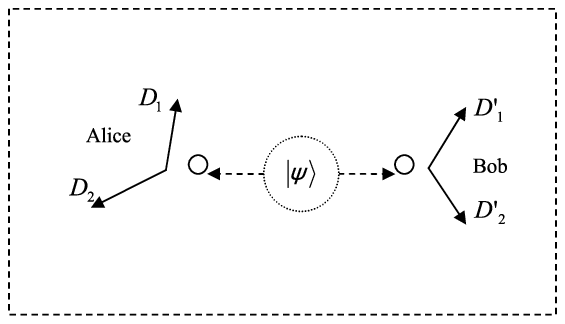';file-properties "XNPEU";}}

As Fig.~2 shows, the two directions refer to two possible directions along
which measurements can be taken and in a run the spin or polarization of the
received half is measured. We call $D_{1}$ and $D_{2}$ observer 1's two
directions and $D_{1}^{\prime }$ and $D_{2}^{\prime }$ observer 2's two
directions. In a run, each observer selects one direction and thus a
directional pair from $(D_{1},$ $D_{2}^{\prime }),$ $(D_{1},D_{1}^{\prime
}), $ $(D_{2},D_{1}^{\prime }),$ $(D_{2},D_{2}^{\prime })$ is selected by
the observers in that run. The Stern-Gerlach type detectors are now rotated
along these selected directions to perform the quantum measurement.
Independent of which directional pair is chosen by the two observers, the
outcome of the quantum measurement is either $+1$ or $-1$ along a
measurement direction.

The relevant 16 probabilities are given \cite{Cereceda} in the following,

\begin{equation}
\begin{array}{c}
\text{Observer }1%
\end{array}%
\begin{array}{c}
\underset{}{%
\begin{array}{c}
D_{1}%
\end{array}%
\begin{array}{c}
+1 \\ 
-1%
\end{array}%
} \\ 
\overset{}{%
\begin{array}{c}
D_{2}%
\end{array}%
\begin{array}{c}
+1 \\ 
-1%
\end{array}%
}%
\end{array}%
\overset{\overset{%
\begin{array}{c}
\text{Observer }2%
\end{array}%
}{%
\begin{array}{cc}
\overset{%
\begin{array}{c}
D_{1}^{\prime }%
\end{array}%
}{%
\begin{array}{cc}
+1 & -1%
\end{array}%
} & \overset{%
\begin{array}{c}
D_{2}^{\prime }%
\end{array}%
}{%
\begin{array}{cc}
+1 & -1%
\end{array}%
}%
\end{array}%
}}{\left( 
\begin{tabular}{c|c}
$\underset{}{%
\begin{array}{cc}
\varepsilon _{1} & \varepsilon _{2} \\ 
\varepsilon _{3} & \varepsilon _{4}%
\end{array}%
}$ & $\underset{}{%
\begin{array}{cc}
\varepsilon _{5} & \varepsilon _{6} \\ 
\varepsilon _{7} & \varepsilon _{8}%
\end{array}%
}$ \\ \hline
$\overset{}{%
\begin{array}{cc}
\varepsilon _{9} & \varepsilon _{10} \\ 
\varepsilon _{11} & \varepsilon _{12}%
\end{array}%
}$ & $\overset{}{%
\begin{array}{cc}
\varepsilon _{13} & \varepsilon _{14} \\ 
\varepsilon _{15} & \varepsilon _{16}%
\end{array}%
}$%
\end{tabular}%
\right) },  \label{table}
\end{equation}%
where, for instance, when the observer $1$ selects the direction $D_{2}$ and
observer $2$ selects the direction $D_{1}^{\prime }$, and the Stern-Gerlach
detectors are rotated along these directions, the probability that both
experimental outcomes are $-1$ is $\varepsilon _{12}$ and the probability
that the observer $1$'s experimental outcome is $+1$ and observer $2$'s
experimental outcome is $-1$ is given by $\varepsilon _{10}$. Appendix A
described how the probabilities $\varepsilon _{j}$ are obtained from a pure
state of two qubits.

We note that there are $16$ probabilities in the above setting of
generalized EPR experiments and that the Fig.~1 giving a normal form
representation of a Bayesian game of incomplete information also has the
same number of entries in it. This naturally leads us to consider the
situation in which the EPR probabilities are taken as the underlying
probabilities of the strategy pairs in the Bayesian game in Fig.~1.

Now, the EPR probabilities can become non-factorizable. As we see in the
following, this consideration motivates us to investigate how the outcome of
the Bayesian game is affected when the underlying probabilities of this game
are obtained from EPR experiments and can thus become non-factorizable.

\subsection{Constraints on EPR probabilities and defining the payoff
relations}

The elements of the probability set $\varepsilon _{j}$ $(1\leq j\leq 16)$
are known to satisfy certain other constraints that are described as
follows. Note that when the directional pair $(D_{1},$ $D_{2}^{\prime })$ is
chosen for all runs of the experiment, the only possible outcomes are $%
(+1,+1),$ $(+1,-1),$ $(-1,+1),$ $(-1,-1)$. The same is true for other
directional pairs $(D_{1},D_{1}^{\prime }),$ $(D_{2},D_{1}^{\prime }),$ $%
(D_{2},D_{2}^{\prime })$. This leads to the normalization constraint:

\begin{equation}
\tsum\nolimits_{j=1}^{4}\varepsilon
_{j}=1=\tsum\nolimits_{j=5}^{8}\varepsilon _{j},\text{ \ \ }%
\tsum\nolimits_{j=9}^{12}\varepsilon
_{j}=1=\tsum\nolimits_{j=13}^{16}\varepsilon _{j}.  \label{normalization}
\end{equation}%
Also, in a particular run of the EPR experiment, the outcome of $+1$ or $-1$
(obtained along the direction $D_{1}$ or direction $D_{2}$) is independent
of whether the direction $D_{1}^{\prime }$ or the direction $D_{2}^{\prime }$
is chosen in that run. Similarly, the outcome of $+1$ or $-1$ (obtained
along $D_{1}^{\prime }$ or $D_{2}^{\prime }$) is independent of whether the
direction $D_{1}$ or the direction $D_{2}$ is chosen in that run. These
requirements, when translated in terms of the probability set $\varepsilon
_{j}$ are expressed as

\begin{eqnarray}
&&%
\begin{array}{cccc}
\varepsilon _{1}+\varepsilon _{2}=\varepsilon _{5}+\varepsilon _{6}, & 
\varepsilon _{1}+\varepsilon _{3}=\varepsilon _{9}+\varepsilon _{11}, & 
\varepsilon _{9}+\varepsilon _{10}=\varepsilon _{13}+\varepsilon _{14}, & 
\varepsilon _{5}+\varepsilon _{7}=\varepsilon _{13}+\varepsilon _{15}, \\ 
\varepsilon _{3}+\varepsilon _{4}=\varepsilon _{7}+\varepsilon _{8}, & 
\varepsilon _{11}+\varepsilon _{12}=\varepsilon _{15}+\varepsilon _{16}, & 
\varepsilon _{2}+\varepsilon _{4}=\varepsilon _{10}+\varepsilon _{12}, & 
\varepsilon _{6}+\varepsilon _{8}=\varepsilon _{14}+\varepsilon _{16}.%
\end{array}
\notag \\
&&  \label{LocalityConstraint}
\end{eqnarray}

A convenient solution of the system (\ref{normalization}, \ref%
{LocalityConstraint}) is reported by Cereceda \cite{Cereceda} to be the one
for which the set of probabilities $\upsilon =\left\{ \varepsilon _{2},\text{
}\varepsilon _{3},\text{ }\varepsilon _{6},\text{ }\varepsilon _{7},\text{ }%
\varepsilon _{10},\text{ }\varepsilon _{11},\text{ }\varepsilon _{13},\text{ 
}\varepsilon _{16}\right\} $ is expressed in terms of the remaining set of
probabilities $\mu =\left\{ \varepsilon _{1},\text{ }\varepsilon _{4},\text{ 
}\varepsilon _{5},\text{ }\varepsilon _{8},\text{ }\varepsilon _{9},\text{ }%
\varepsilon _{12},\text{ }\varepsilon _{14},\text{ }\varepsilon
_{15}\right\} $ that is given as

\begin{equation}
\begin{array}{l}
\varepsilon _{2}=(1-\varepsilon _{1}-\varepsilon _{4}+\varepsilon
_{5}-\varepsilon _{8}-\varepsilon _{9}+\varepsilon _{12}+\varepsilon
_{14}-\varepsilon _{15})/2, \\ 
\varepsilon _{3}=(1-\varepsilon _{1}-\varepsilon _{4}-\varepsilon
_{5}+\varepsilon _{8}+\varepsilon _{9}-\varepsilon _{12}-\varepsilon
_{14}+\varepsilon _{15})/2, \\ 
\varepsilon _{6}=(1+\varepsilon _{1}-\varepsilon _{4}-\varepsilon
_{5}-\varepsilon _{8}-\varepsilon _{9}+\varepsilon _{12}+\varepsilon
_{14}-\varepsilon _{15})/2, \\ 
\varepsilon _{7}=(1-\varepsilon _{1}+\varepsilon _{4}-\varepsilon
_{5}-\varepsilon _{8}+\varepsilon _{9}-\varepsilon _{12}-\varepsilon
_{14}+\varepsilon _{15})/2, \\ 
\varepsilon _{10}=(1-\varepsilon _{1}+\varepsilon _{4}+\varepsilon
_{5}-\varepsilon _{8}-\varepsilon _{9}-\varepsilon _{12}+\varepsilon
_{14}-\varepsilon _{15})/2, \\ 
\varepsilon _{11}=(1+\varepsilon _{1}-\varepsilon _{4}-\varepsilon
_{5}+\varepsilon _{8}-\varepsilon _{9}-\varepsilon _{12}-\varepsilon
_{14}+\varepsilon _{15})/2, \\ 
\varepsilon _{13}=(1-\varepsilon _{1}+\varepsilon _{4}+\varepsilon
_{5}-\varepsilon _{8}+\varepsilon _{9}-\varepsilon _{12}-\varepsilon
_{14}-\varepsilon _{15})/2, \\ 
\varepsilon _{16}=(1+\varepsilon _{1}-\varepsilon _{4}-\varepsilon
_{5}+\varepsilon _{8}-\varepsilon _{9}+\varepsilon _{12}-\varepsilon
_{14}-\varepsilon _{15})/2.%
\end{array}
\label{dependentProbabilities}
\end{equation}%
This allows us to consider the elements of the set $\mu $ as independent
variables.

In order to use the EPR setting to play the Bayesian game in Fig.~1, we call
the observers $1$ and $2$ the players Alice and Bob, respectively, of the
Bayesian game. We then associate one half of the EPR pair to the player
Alice and the other half to the player Bob. As Alice and Bob have two
directions each, we associate Alice's two directions to the two types of
Alice. That is $D_{1}\sim $Alice's type 1 and $D_{2}\sim $Alice's type $2$.
Similarly, we associate Bob's two directions to the two types of Bob. That
is $D_{1}^{\prime }\sim $Bob's type 1 and $D_{2}^{\prime }\sim $Bob's type 2.

With these associations, and in view of the game in Fig.~1, in a run, each
of the two directions $D_{1}$ and $D_{2}$ is chosen with probability $\frac{1%
}{2}$. Similarly, the directions $D_{1}^{\prime }$ and $D_{2}^{\prime }$ are
chosen with probabilities $\omega $ and $(1-\omega )$, respectively. In view
of the payoff relations (\ref{MixedStrategyPayoffs}) in the factorizable
game, the players' payoff relations in the game with EPR probabilities can
now be expressed as

\begin{eqnarray}
\Pi _{\text{A}_{1}}(\varepsilon _{j}) &=&\frac{1}{2}\left\{ (2)\varepsilon
_{1}+(0)\varepsilon _{2}+(0)\varepsilon _{3}+(1)\varepsilon _{4}\right\} +%
\frac{1}{2}\left\{ (2)\varepsilon _{5}+(0)\varepsilon _{6}+(0)\varepsilon
_{7}+(1)\varepsilon _{8}\right\} ,  \notag \\
\Pi _{\text{A}_{2}}(\varepsilon _{j}) &=&\frac{1}{2}\left\{ (0)\varepsilon
_{9}+(2)\varepsilon _{10}+(1)\varepsilon _{11}+(0)\varepsilon _{12}\right\} +%
\frac{1}{2}\left\{ (0)\varepsilon _{13}+(2)\varepsilon _{14}+(1)\varepsilon
_{15}+(0)\varepsilon _{16}\right\} ,  \notag \\
\Pi _{\text{B}_{1}}(\varepsilon _{j}) &=&\omega \left\{ (1)\varepsilon
_{1}+(0)\varepsilon _{2}+(0)\varepsilon _{3}+(2)\varepsilon _{4}\right\}
+(1-\omega )\left\{ (1)\varepsilon _{9}+(0)\varepsilon _{10}+(0)\varepsilon
_{11}+(2)\varepsilon _{12}\right\} ,  \notag \\
\Pi _{\text{B}_{2}}(\varepsilon _{j}) &=&\omega \left\{ (0)\varepsilon
_{5}+(2)\varepsilon _{6}+(1)\varepsilon _{7}+(0)\varepsilon _{8}\right\}
+(1-\omega )\left\{ (0)\varepsilon _{13}+(2)\varepsilon _{14}+(1)\varepsilon
_{15}+(0)\varepsilon _{16}\right\} ,  \notag \\
&&  \label{EPRpayoffs}
\end{eqnarray}%
where $1\leq j\leq 16$ and $\Pi _{\text{A}_{1}}(\varepsilon _{j}),$ $\Pi _{%
\text{A}_{2}}(\varepsilon _{j}),$ $\Pi _{\text{B}_{1}}(\varepsilon _{j}),$
and $\Pi _{\text{B}_{2}}(\varepsilon _{j})$ are the payoffs to Alice of type 
$1$, Alice of type $2$, Bob of type $1$, and Bob of type $2$, respectively,
expressed in term of the probabilities $\varepsilon _{j}$ that are defined
in (\ref{table}).

Note that the payoffs (\ref{EPRpayoffs}) are reduced to the payoffs in the
mixed-strategy game (\ref{MixedStrategyPayoffs}) when the probability set $%
\varepsilon _{j}$ $(1\leq j\leq 16)$ is factorizable in terms of the
probabilities $p,$ $q,$ $p^{\prime },$ $q^{\prime }\in \lbrack 0,1]$ as
given by Eqs.~(\ref{factorizability}).

\section{Nash equilibrium inequalities in Bayesian game of Battle of Sexes
with EPR probabilities}

So as to find the Nash equilibria when the probability set $\varepsilon _{j}$
$(1\leq j\leq 16)$ are non-factorizable, we notice that when $\varepsilon
_{j}$ are factorizable in terms of the probabilities $p,$ $q,$ $p^{\prime },$
and $q^{\prime }$, we can write

\begin{align}
\varepsilon _{1}& =pp^{\prime },\text{ }\varepsilon _{2}=p(1-p^{\prime }),%
\text{ }\varepsilon _{3}=(1-p)p^{\prime },\text{ }\varepsilon
_{4}=(1-p)(1-p^{\prime }),  \notag \\
\varepsilon _{5}& =pq^{\prime },\text{ }\varepsilon _{6}=p(1-q^{\prime }),%
\text{ }\varepsilon _{7}=(1-p)q^{\prime },\text{ }\varepsilon
_{8}=(1-p)(1-q^{\prime }),  \notag \\
\varepsilon _{9}& =qp^{\prime },\text{ }\varepsilon _{10}=q(1-p^{\prime }),%
\text{ }\varepsilon _{11}=(1-q)p^{\prime },\text{ }\varepsilon
_{12}=(1-q)(1-p^{\prime }),  \notag \\
\varepsilon _{13}& =qq^{\prime },\text{ }\varepsilon _{14}=q(1-q^{\prime }),%
\text{ }\varepsilon _{15}=(1-q)q^{\prime },\text{ }\varepsilon
_{16}=(1-q)(1-q^{\prime }).  \label{factorizability}
\end{align}%
With this the payoff relations (\ref{EPRpayoffs}) are reduced to the payoffs
given by (\ref{MixedStrategyPayoffs}). Also, when $\varepsilon _{j}$ are
factorizable, and can be expressed in terms of $p,$ $q,$ $p^{\prime },$ and $%
q^{\prime }$, they satisfy the constraints given by Eqs.~(\ref{normalization}%
, \ref{LocalityConstraint}) and $p,$ $q,$ $p^{\prime },$ and $q^{\prime }$
can then be expressed in terms of the probabilities $\varepsilon _{j}$ as
follows

\begin{eqnarray}
p &=&\frac{1}{2}(\varepsilon _{1}+\varepsilon _{2}+\varepsilon
_{5}+\varepsilon _{6}),\text{ }q=\frac{1}{2}(\varepsilon _{9}+\varepsilon
_{10}+\varepsilon _{13}+\varepsilon _{14}),  \notag \\
p^{\prime } &=&\frac{1}{2}(\varepsilon _{1}+\varepsilon _{3}+\varepsilon
_{9}+\varepsilon _{11}),\text{ }q^{\prime }=\frac{1}{2}(\varepsilon
_{5}+\varepsilon _{7}+\varepsilon _{13}+\varepsilon _{15}).
\label{re-expressedProbs}
\end{eqnarray}%
From the set of inequalities (\ref{NashEquilibria}), the expressions
describing the Nash equilibria in the factorizable game are

\begin{gather}
\frac{\partial \Pi _{\text{A}_{1}}}{\partial p}\mid _{\ast }(p^{\ast
}-p)\geq 0,\text{ }\frac{\partial \Pi _{\text{A}_{2}}}{\partial q}\mid
_{\ast }(q^{\ast }-q)\geq 0,  \notag \\
\frac{\partial \Pi _{\text{B}_{1}}}{\partial p^{\prime }}\mid _{\ast
}(p^{\prime \ast }-p^{\prime })\geq 0,\text{ }\frac{\partial \Pi _{\text{B}%
_{2}}}{\partial q^{\prime }}\mid _{\ast }(q^{\prime \ast }-q^{\prime })\geq
0.  \label{Nash}
\end{gather}

To evaluate these inequalities, we use the relations (\ref%
{dependentProbabilities}) to express the payoff relations (\ref{EPRpayoffs})
in terms of the elements from the set $\mu $ as follows

\begin{eqnarray}
\Pi _{\text{A}_{1}}(\varepsilon _{j}) &=&\frac{1}{2}(2\varepsilon
_{1}+\varepsilon _{4}+2\varepsilon _{5}+\varepsilon _{8}),\text{ }  \notag \\
\Pi _{\text{A}_{2}}(\varepsilon _{j}) &=&\frac{1}{4}(3-\varepsilon
_{1}+\varepsilon _{4}+\varepsilon _{5}-\varepsilon _{8}-3\varepsilon
_{9}-3\varepsilon _{12}+5\varepsilon _{14}+\varepsilon _{15}),  \notag \\
\Pi _{\text{B}_{1}}(\varepsilon _{j}) &=&\omega (\varepsilon
_{1}+2\varepsilon _{4}-\varepsilon _{9}-2\varepsilon _{12})+(\varepsilon
_{9}+2\varepsilon _{12})  \notag \\
\Pi _{\text{B}_{2}}(\varepsilon _{j}) &=&\frac{\omega }{2}(3+\varepsilon
_{1}-\varepsilon _{4}-3\varepsilon _{5}-3\varepsilon _{8}-\varepsilon
_{9}+\varepsilon _{12}-3\varepsilon _{14}-3\varepsilon _{15})+(2\varepsilon
_{14}+\varepsilon _{15}).  \label{Players' Payoffs}
\end{eqnarray}%
Similarly, using the relations (\ref{dependentProbabilities}), we express $%
p, $ $q,$ $p^{\prime },$ $q^{\prime }$ in (\ref{re-expressedProbs}) in terms
of the elements from the set $\mu $ as

\begin{eqnarray}
p &=&\frac{1}{2}(1+\varepsilon _{1}-\varepsilon _{4}+\varepsilon
_{5}-\varepsilon _{8}-\varepsilon _{9}+\varepsilon _{12}+\varepsilon
_{14}-\varepsilon _{15}),  \notag \\
q &=&\frac{1}{2}(1-\varepsilon _{1}+\varepsilon _{4}+\varepsilon
_{5}-\varepsilon _{8}+\varepsilon _{9}-\varepsilon _{12}+\varepsilon
_{14}-\varepsilon _{15}),  \notag \\
p^{\prime } &=&\frac{1}{2}(1+\varepsilon _{1}-\varepsilon _{4}-\varepsilon
_{5}+\varepsilon _{8}+\varepsilon _{9}-\varepsilon _{12}-\varepsilon
_{14}+\varepsilon _{15}),  \notag \\
q^{\prime } &=&\frac{1}{2}(1-\varepsilon _{1}+\varepsilon _{4}+\varepsilon
_{5}-\varepsilon _{8}+\varepsilon _{9}-\varepsilon _{12}-\varepsilon
_{14}+\varepsilon _{15}).  \label{dependents}
\end{eqnarray}%
This allows us to use the chain rule to evaluate

\begin{eqnarray}
\frac{\partial \Pi _{\text{A}_{1}}(\varepsilon _{j})}{\partial p} &=&\frac{%
\partial \Pi _{\text{A}_{1}}(\varepsilon _{j})}{\partial \varepsilon _{1}}%
\frac{\partial \varepsilon _{1}}{\partial p}+\frac{\partial \Pi _{\text{A}%
_{1}}(\varepsilon _{j})}{\partial \varepsilon _{4}}\frac{\partial
\varepsilon _{4}}{\partial p}+\frac{\partial \Pi _{\text{A}_{1}}(\varepsilon
_{j})}{\partial \varepsilon _{5}}\frac{\partial \varepsilon _{5}}{\partial p}%
+\frac{\partial \Pi _{\text{A}_{1}}(\varepsilon _{j})}{\partial \varepsilon
_{8}}\frac{\partial \varepsilon _{8}}{\partial p}+  \notag \\
&&\frac{\partial \Pi _{\text{A}_{1}}(\varepsilon _{j})}{\partial \varepsilon
_{9}}\frac{\partial \varepsilon _{9}}{\partial p}+\frac{\partial \Pi _{\text{%
A}_{1}}(\varepsilon _{j})}{\partial \varepsilon _{12}}\frac{\partial
\varepsilon _{12}}{\partial p}+\frac{\partial \Pi _{\text{A}%
_{1}}(\varepsilon _{j})}{\partial \varepsilon _{14}}\frac{\partial
\varepsilon _{14}}{\partial p}+\frac{\partial \Pi _{\text{A}%
_{1}}(\varepsilon _{j})}{\partial \varepsilon _{15}}\frac{\partial
\varepsilon _{15}}{\partial p},  \notag \\
&&
\end{eqnarray}%
that gives $\frac{\partial \Pi _{\text{A}_{1}}(\varepsilon _{j})}{\partial p}%
=2$. Similarly, we obtain

\begin{equation}
\frac{\partial \Pi _{\text{A}_{2}}(\varepsilon _{j})}{\partial q}=4,\text{ }%
\frac{\partial \Pi _{\text{B}_{1}}(\varepsilon _{j})}{\partial p^{\prime }}%
=-2,\text{ }\frac{\partial \Pi _{\text{B}_{2}}(\varepsilon _{j})}{\partial
q^{\prime }}=-2(2\omega +1).
\end{equation}%
The inequalities (\ref{Nash}) are now written as

\begin{gather}
2(p^{\ast }-p)\geq 0,\text{ }4(q^{\ast }-q)\geq 0,  \notag \\
-2(p^{\prime \ast }-p^{\prime })\geq 0,\text{ }-2(2\omega +1)(q^{\prime \ast
}-q^{\prime })\geq 0,  \label{NEConditions}
\end{gather}%
giving

\begin{eqnarray}
p^{\ast } &=&\frac{1}{2}(\varepsilon _{1}^{\ast }+\varepsilon _{2}^{\ast
}+\varepsilon _{5}^{\ast }+\varepsilon _{6}^{\ast })=1,\text{ }q^{\ast }=%
\frac{1}{2}(\varepsilon _{9}^{\ast }+\varepsilon _{10}^{\ast }+\varepsilon
_{13}^{\ast }+\varepsilon _{14}^{\ast })=1,  \notag \\
p^{\prime \ast } &=&\frac{1}{2}(\varepsilon _{1}^{\ast }+\varepsilon
_{3}^{\ast }+\varepsilon _{9}^{\ast }+\varepsilon _{11}^{\ast })=0,\text{ }%
q^{\prime \ast }=\frac{1}{2}(\varepsilon _{5}^{\ast }+\varepsilon _{7}^{\ast
}+\varepsilon _{13}^{\ast }+\varepsilon _{15}^{\ast })=0,  \label{solutions}
\end{eqnarray}%
as a unique Bayesian Nash equilibrium of the game. Using Eqs.~(\ref%
{re-expressedProbs},\ref{dependents}) we then obtain $\varepsilon _{2}^{\ast
}=\varepsilon _{6}^{\ast }=\varepsilon _{10}^{\ast }=\varepsilon _{14}^{\ast
}=1$ with all the rest of $\varepsilon _{j}^{\ast }$ being zeros. The
players' payoffs at this equilibrium are obtained as

\begin{equation}
\Pi _{\text{A}_{1}}(\varepsilon _{j}^{\ast })=0,\text{ }\Pi _{\text{A}%
_{2}}(\varepsilon _{j}^{\ast })=2,\text{ }\Pi _{\text{B}_{1}}(\varepsilon
_{j}^{\ast })=0,\text{ }\Pi _{\text{B}_{2}}(\varepsilon _{j}^{\ast })=2,
\label{SolutionPayoffs}
\end{equation}%
presenting a dramatic contrast to what is the case when the underlying
probabilities are factorizable.

\section{Discussion}

As is well known, Bell's inequality presents stark difference between the
classical and quantum world. The CHSH form of Bell's inequality can be
expressed as a constraint on probabilities without referring to the
formalism and mathematical machinery of quantum mechanics. Game theory is
based on the theory of probability and this suggests that the quantum
contents of quantum games can be given an expression that only employ
probabilities without referring to the formalism of quantum mechanics. A
probabilistic approach to quantum games is only a matter of perspective as
the quantum Bayesian game discussed in this paper can be physically
implemented using EPR type experiments that result in non-factorizable
probabilities.

Essentially, this paper discusses a physical implementation of a quantum
Bayesian game using EPR experiments. We find that these experiments provide
a natural setting for analyzing a quantum Bayesian game. Our analysis uses
only probabilities as they facilitate wider access to the area of quantum
games. The physical realization of our game is provided by actual EPR
experiments where quantum mechanics makes the difference and which can also
be expressed in terms of probabilities only.

We study the game of Battle of Sexes with incomplete information that has
both pure and mixed Bayesian Nash equilibria. We investigate the situation
when the underlying probabilities of this game can become non-factorizable.
As is known, the probabilities in generalized EPR experiments can become
non-factorizable and in certain cases can maximally violate the
corresponding CHSH version of Bell's inequality. When the quantum mechanical
probabilities are factorizable the game attains a classical interpretation.
However, when the probabilities are allowed to become non-factorizable, we
find that the solution of the game turns out to be entirely different and,
in contrast to the classical game that has both pure and mixed Bayesian Nash
equilibria, the quantum game has a unique Bayesian Nash equilibrium.

A natural question to ask here is whether any set of probabilities that
satisfies the constraints (\ref{normalization},\ref{LocalityConstraint}) is
physically realizable? Quantum mechanics is known to impose further
constraints and one such constraint is given by the CHSH version of Bell's
inequality \cite{CHSH,Peres,Cereceda}. This constraint states that the
quantity $\Delta $ defined by

\begin{equation}
\Delta =2(\varepsilon _{1}+\varepsilon _{4}+\varepsilon _{5}+\varepsilon
_{8}+\varepsilon _{9}+\varepsilon _{12}+\varepsilon _{14}+\varepsilon
_{15}-2)  \label{CHSH delta}
\end{equation}%
is restricted in the range $\left\vert \Delta _{QM}\right\vert \leq 2\sqrt{2}
$ by the laws of quantum mechanics and the CHSH version of Bell's inequality
is violated in quantum mechanical experiments when $2<\left\vert \Delta
\right\vert $. These constraints are imposed by physical realizations and a
maximum value of $\Delta =4$ emerges when only non-negative probabilities
are considered. It is known that this value is not physically realizable.

Our analysis for a Bayesian game of Battle of Sexes shows that when the
underlying probabilities are obtained from generalized EPR experiments, and
thus can be non-factorizable, a unique Bayesian Nash equilibrium of the game
emerges. This equilibrium corresponds to $\varepsilon _{2}^{\ast
}=\varepsilon _{6}^{\ast }=\varepsilon _{10}^{\ast }=\varepsilon _{14}^{\ast
}=1$ with the remaining $\varepsilon _{j}^{\ast }$ being zeros. Substituting
these values in Eq.~(\ref{CHSH delta}) gives $\Delta =-2$ i.e. the unique
Bayesian Nash equilibrium is obtained without violating the CHSH version of
Bell's inequality.

A particularly interesting situation would be when the violation of CHSH
version of Bell's inequality leads to a new outcome of the game. Quantum
mechanics is known to put its own constraints on the allowed ranges of the
variables $(\varepsilon _{1}+\varepsilon _{2}+\varepsilon _{5}+\varepsilon
_{6}),$ $(\varepsilon _{9}+\varepsilon _{10}+\varepsilon _{13}+\varepsilon
_{14}),$ $(\varepsilon _{1}+\varepsilon _{3}+\varepsilon _{9}+\varepsilon
_{11}),$ and $(\varepsilon _{5}+\varepsilon _{7}+\varepsilon
_{13}+\varepsilon _{15})$, as described in Eqs.~(\ref{re-expressedProbs}),
and they are different from those imposed by just not permitting
probabilities to have negative values. An important question would then be
to ask whether this would change or affect the outcome of the game. A
consideration of Eqs.~(\ref{NEConditions}, \ref{solutions}) shows that the
conditions yielding the Bayesian Nash equilibrium are simply too strong to
be affected by extra constraints that quantum mechanics can impose on the
allowed ranges of the variables $(\varepsilon _{1}+\varepsilon
_{2}+\varepsilon _{5}+\varepsilon _{6}),$ $(\varepsilon _{9}+\varepsilon
_{10}+\varepsilon _{13}+\varepsilon _{14}),$ $(\varepsilon _{1}+\varepsilon
_{3}+\varepsilon _{9}+\varepsilon _{11}),$ and $(\varepsilon
_{5}+\varepsilon _{7}+\varepsilon _{13}+\varepsilon _{15})$. However, it is
possible that for other Bayesian games the conditions giving the outcome are
not so strong and it then would be worthwhile to investigate this question
further.

\textbf{Acknowledgement}: We acknowledge helpful discussions with Andrew
Allison.

\section{Appendix A}

Using the standard notation, the set of 16 probabilities in (\ref{table}) is
obtained from a pure state of two qubits

\begin{equation}
\left\vert \psi _{0}\right\rangle =\alpha \left\vert 00\right\rangle +\beta
\left\vert 01\right\rangle +\gamma \left\vert 10\right\rangle +\delta
\left\vert 11\right\rangle  \label{2-qubit pure state}
\end{equation}%
where $\alpha ,\beta ,\gamma ,\delta \in \boldsymbol{C}$ and $\left\vert
\alpha \right\vert ^{2}+\left\vert \beta \right\vert ^{2}+\left\vert \gamma
\right\vert ^{2}+\left\vert \delta \right\vert ^{2}=1$. We assume that
observer $1$'s directions $D_{1}$ and $D_{2}$ are along the unit vectors $%
\hat{a}$ and $\hat{c}$, respectively. Similarly, observer $2$'s directions $%
D_{1}^{\prime }$ and $D_{2}^{\prime }$ are along the unit vectors $\hat{b}$
and $\hat{d}$, respectively. Without loss of generality, we also assume that
the unit vectors $\hat{a}=[a_{x},a_{y}]$, $\hat{b}=[b_{x},b_{y}]$, $\hat{c}%
=[c_{x},c_{y}]$, and $\hat{d}=[d_{x},d_{y}]$ are all located in the x-y
plane. Observer $1$'s measurement operators are then $\hat{\sigma}\cdot \hat{%
a}$ and $\hat{\sigma}\cdot \hat{c}$, respectively. Similarly, Observer $2$'s
measurement operators are $\hat{\sigma}\cdot \hat{b}$ and $\hat{\sigma}\cdot 
\hat{d}$, respectively. Here $\hat{\sigma}=[\sigma _{x},\sigma _{y},\sigma
_{z}]$ and $\sigma _{x},\sigma _{y},\sigma _{z}$ are Pauli spin matrices.

Consider the probability $\varepsilon _{1}$ in the (\ref{table}) that
corresponds to observers $1$ and $2$ measuring along the directions $\hat{a}$
and $\hat{b}$, respectively, and both obtaining the outcome $+1$. For this
situation, we require the eigenstates of the operators $(\hat{\sigma}\cdot 
\hat{a})$ and $(\hat{\sigma}\cdot \hat{b})$, with the eigenvalue of $+1$ for
both. These are found to be $\frac{\left\vert 0\right\rangle
+(a_{x}+ia_{y})\left\vert 1\right\rangle }{\sqrt{2}}$ and $\frac{\left\vert
0\right\rangle +(b_{x}+ib_{y})\left\vert 1\right\rangle }{\sqrt{2}}$,
respectively. From these the eigenstate of the measurement operator $(\hat{%
\sigma}\cdot \hat{a})\otimes (\hat{\sigma}\cdot \hat{b})$, with the
eigenvalue $+1$, is obtained as

\begin{equation}
\left\vert \psi _{1}\right\rangle =\frac{1}{2}(\left\vert 00\right\rangle
+(b_{x}+ib_{y})\left\vert 01\right\rangle +(a_{x}+ia_{y})\left\vert
10\right\rangle +(a_{x}+ia_{y})(b_{x}+ib_{y})\left\vert 11\right\rangle ),
\end{equation}%
and the probability $\varepsilon _{1}$ is then obtained from $\left\vert
\left\langle \psi _{1}\right\vert \left. \psi _{0}\right\rangle \right\vert
^{2}$. For the pure state (\ref{2-qubit pure state}) this becomes

\begin{equation}
\varepsilon _{1}=\frac{1}{4}\left\vert \alpha +\beta (b_{x}-ib_{y})+\gamma
(a_{x}-ia_{y})+\delta (a_{x}-ia_{y})(b_{x}-ib_{y})\right\vert ^{2}.
\end{equation}%
Similarly, for the probability $\varepsilon _{2}$, along with the eigenstate
of $(\hat{\sigma}\cdot \hat{a})$ with eigenvalue $+1$ obtained above, we
require the eigenstate of the operator $(\hat{\sigma}\cdot \hat{b})$ with
the eigenvalues $-1$, which is $\frac{\left\vert 0\right\rangle
-(b_{x}+ib_{y})\left\vert 1\right\rangle }{\sqrt{2}}$. From these the
eigenstate of the measurement operator $(\hat{\sigma}\cdot \hat{a})\otimes (%
\hat{\sigma}\cdot \hat{b})$, with the eigenvalue $-1$, is obtained as

\begin{equation}
\left\vert \psi _{2}\right\rangle =\frac{1}{2}(\left\vert 00\right\rangle
-(b_{x}+ib_{y})\left\vert 01\right\rangle +(a_{x}+ia_{y})\left\vert
10\right\rangle -(a_{x}+ia_{y})(b_{x}+ib_{y})\left\vert 11\right\rangle ),
\end{equation}%
and, as before, the probability $\varepsilon _{2}$ is then obtained from $%
\left\vert \left\langle \psi _{2}\right\vert \left. \psi _{0}\right\rangle
\right\vert ^{2}$. For the pure state (\ref{2-qubit pure state}) this becomes

\begin{equation}
\varepsilon _{2}=\frac{1}{4}\left\vert \alpha -\beta (b_{x}-ib_{y})+\gamma
(a_{x}-ia_{y})-\delta (a_{x}-ia_{y})(b_{x}-ib_{y})\right\vert ^{2},
\end{equation}%
and the remaining probabilities $\varepsilon _{3},\varepsilon
_{4}...\varepsilon _{16}$ can be obtained similarly.

\end{document}